\documentstyle{article}

\begin{document}

\title{ $SU(3) \times SU(2) \times U(1):$  The residual symmetry of extended
conformal gravity.}
\author{James T. Wheeler \\ {\em Utah State University, Logan, UT 84322}}
\maketitle

\section{Motivation}

There are two points crucial to the understanding of gravity and its
relationship to the standard model of the remaining forces \cite{Stdmodel}.
These observations motivate the subsequent search for an additional symmetry
and
ultimately lead to a natural extension of the conformal algebra.

     \begin{enumerate}
         \item Conformal symmetry must underlie our mathematical
                    description of nature.
         \item Naive conformal gravity fails.
     \end{enumerate}

The first claim is true because we can measure only relative magnitudes.
When the size of a table is given in meters it is being compared to a certain
number of wavelengths of light;  the mass of a chair is ultimately in
comparison
to the masses of particles or nuclei.  The situation is no different at the
subatomic level, where we maintain our use of standard units.

Nor is it necessary for our standard units to be chosen in the same way
everywhere.  The choice is made in accordance with the purpose at hand.  A
physicist generally chooses a uniform scale because with such a scale motion
has
a simple description.  But an artist represents objects using a scale having
distant objects smaller than the same types of object nearby because this is
consistent with our visual process.  It is certainly possible and consistent to
translate our physical descriptions into such units.

While the Poincar\'{e} group does not generate scale changes, the conformal
group respects this freedom of choosing scale by preserving only relative
lengths.  The preservation of one scale choice by the Poincar\'{e} group is not
physical -- scale need not be preserved.  Thus, we reach the first conclusion:
the local spacetime symmetry which does not claim more than we can know is the
conformal group, not the Poincar\'{e} group.

To understand the second conclusion requires some understanding of the
conformal
group.  We have, of course, the usual Poincar\'{e} symmetries -- Lorentz
transformations and translations.  Lorentz transformations $(o(3,1))$ or
$sl(2,C))$ include both rotations and boosts, and are generated by $4 \times 4$
matrices, $M_{ab}$.  The translations, with generators $P_{a}$, give simple
displacements.
In addition to these are two new transformations:

         \begin{enumerate}
              \item Scale changes, generated by the dilation operator, $D$
directly produce the allowed choices of units by rescaling lengths locally.
              \item Conformal translations.  The four generators, $K_{a}$,
translate infinitesimal vectors in inverse coordinates.  Such a vector is
inverted through a unit sphere, translated, then re-inverted.  The
transformation clearly involves an arbitrary displacement vector.  The thing to
notice -- or rather not to notice -- is that the vector is also rotated and
rescaled.  Since both rotating and stretching are already independent
transformations, the group closes.
          \end{enumerate}

It is central to what follows to understand that with the scale symmetry
available, one mode of translation is just as good as the other.  Since we only
compare relative magnitude, translating with $K_{a}$ or $P_{a}$ always gives
the
same physical result:  things are moved from here to there and have the same
relative magnitude.
We now return to the second motivational observation.  Seeking a conformal
theory of gravity, we introduce gauge fields for each of the generators of the
algebra:

     \[ M_{ab}, P_{a}, K_{a}, D   \longrightarrow  \omega_{\alpha}^{ab},
u_{\alpha}^a, v_{\alpha}^a, W_{\alpha} \]

When we gauge the Poincar\'{e} group, the first of these,
$\omega_{\alpha}^{ab}$, is the spin connection, while the second, the gauge
field of translations, $u_{\alpha}^a,$ is identified with the vierbein,
$e_{\alpha}^a$.  The vierbein (the soldering form of the fibre bundle) is
related to the metric tensor, $g_{\alpha\beta}$, via the relation
                \[g_{\alpha\beta} = e_{\alpha}^a e_{\beta}^b \eta_{ab}\]
where $\eta_{ab}$ is the Minkowski metric.

If we compute the conformal curvature tensor we find that it has certain
components corresponding to a generalization of the usual Riemannian curvature,
and additional components related to the $u_{\alpha}^a,$ $v_{\alpha}^a,$ and
$W_{\alpha}$ fields.
 A problem arises as soon as we write down a scale-invariant
action from this curvature.  It has been shown \cite{Wh1,Kaku} that for any
such
action we write, the $v_{\alpha}^a$ field is auxiliary.  What happens is that,
when the action (which contains 9 independent terms) is varied with respect to
$v_{\alpha}^a,$ the resulting field equation for $v_{\alpha}^a$ can be solved
and substituted back into the action.  The resulting effective action is
independent of $v_{\alpha}^a$.
A second problem arises as soon as we consider the quantization of a
scale-invariant theory.  The action of a dilation on the generator of
translations is
      \[ e^{-\lambda D} P_{a} e^{\lambda D} = e^{\lambda} P_{a} \]
which simply says that the action of a dilation on an infinitesimal translation
is to change the magnitude, but not the direction, of the displacement.  But
since $P^{2}$ is identified with the square of the mass,
this implies the relation
      \[ e^{-\lambda D} P^{2} e^{\lambda D} = e^{2\lambda} P^{2}. \]
As a consequence, masses may be changed by an arbitrary factor, $e^{2\lambda}$.
The mass spectrum becomes continuous instead of discrete, in contradiction with
experiment.  For this reason, it is generally assumed that scale-invariance is
not a good symmetry at the quantum level \cite{Wess,Carruthers}.

Thus, naive implementation of the conformal group leads to the conclusion that
the gauge field of the conformal translations may be eliminated and the
dilation
symmetry must be broken.  We are left with the Poincar\'{e} symmetry from which
we started.

\section{The metric structure of conformal gauge theory}

The conclusions above concerning the elimination of the gauge field,
$v_{\alpha}^a$, hold because we chose to identify the gauge field of
translation, $u_{\alpha}^a$, with the vierbein, and hence with the metric.  Had
we chosen  $v_{\alpha}^a$ as the vierbein instead, it would have been
$u_{\alpha}^a$ that was auxiliary.  The only reason for picking $u_{\alpha}^a$
is that that is what is done when gauging the Poincar\'{e} group.  With the
conformal group, there is an option.
This observation is central.  It means that there is an additional symmetry,
implicit in the conformal group, which did not need to be broken.  The only
thing that breaks the symmetry between the two translations, $P_{a}$
 and $K_{a},$ is
our arbitrary choice.  If we can rewrite or alter the conformal group in such a
way as to make this symmetry explicit, then we may find some new content to
conformal gauge theory after all.

This reworking of the group is dependent on the introduction of a metric  on
the
underlying manifold.  What we're going to do is to find all metrics which can
be
constructed from the gauge fields $u_{\alpha}^a$ and $v_{\alpha}^a$, and
rebuild
the group in a way that guarantees that we can independently pick any of the
possibilities as a gauge choice.

It is easy to establish that there are precisely three rank-2, symmetric tensor
fields constructible from the gauge fields $u_{\alpha}^a$ and $v_{\alpha}^a$.
They are:

        \[ g^{1}_{\alpha\beta} = u_{\alpha}^a u_{\beta}^b \eta_{ab} \]
        \[ g^{2}_{\alpha\beta} = u_{(\alpha}^a v_{\beta)}^b \eta_{ab}  \]
        \[ g^{3}_{\alpha\beta} = v_{\alpha}^a v_{\beta}^b \eta_{ab} \]

While there is no guarantee that any of these metrics is invertible or
torsion-free, the same is true of the gauge theory of the Poincar\'{e} group.
Invertibility and vanishing torsion are assumptions which must be made to
reproduce general relativity from the gauge theory, and we make the same
assumptions here.  The status of these assumptions is a subject of debate.
Certainly, invertibility holds generically.  As for the vanishing of the
torsion, it is still an open question whether torsion does vanish.  While the
macroscopic limits are quite stringent, there is always the possibility of
consistently interpreting some physical field as torsion.

The next step is to write the symmetry so that $g^{1}_{\alpha\beta}$,
$g^{2}_{\alpha\beta}$, and $g^{3}_{\alpha\beta}$ are possible gauge choices.
This is achieved by introducing a vierbein, $e^{i}_{\alpha}{}^{a}  (i = 1, 2,
3)$,
for each possible metric, and introducing a translation generator, $T^{i}_{a}$,
as the operator for which $e^{i}_{\alpha}{}^{a}$ is the gauge field.  Clearly, we
can let $T^{1}_{a} = P_{a}$ and $T^{3}_{a} = K_{a}$, but $T^{2}_{a}$ is new.
One might think that $e^{2}_{\alpha}{}^{a}$ could be gotten by taking some
combination $\lambda_{1}P_{a} + \lambda_{2}K_{a}$, but such combinations always
introduce some measure of $g^{1}_{\alpha\beta}$ and $g^{3}_{\alpha\beta}$ in
addition, so  the middle metric would not be independent.

The presence of the new generator $T^{2}_{a}$ changes the conformal algebra,
and
we are faced with
a choice of  several ways to close the new algebra.  We could simply let
the new translation commute with all of the other generators, but this means
that it rotates as four scalars instead of as a vector.  We can extend the
group
until it can contain the generator we want, or we can contract the group.  Both
of these latter ways work.  For example, $O(5,3)$ contains two translations and
gle of 4-spinor.  Because the norm is preserved,
transformations of $\psi$ produce rotations of $g^{A}.$

When a particular metric is singled out, the $SU(4)$ gauge is partially fixed.
There remains an $SU(3)$ subgroup which leaves $\psi$ invariant.  In addition,
there additional symmetries present.  These symmetries may depend on the
particular $\psi$ chosen.  We prove the following theorem:
\newtheorem{theorem}{Thm.}
\begin{theorem}
Fixing the metric reduces the $SU(4)$ symmetry to $SU(3) \times C(1) \times K,$
where $C(1)$ is a bounded, one-parameter group and $K$ is a discrete group.
When
the spacetime metric has definite scaling weight, $C(1) = U(1)$ and $K$
includes
the integers.

Proof:     We first demonstrate that the subgroup which leaves a fixed spinor
invariant is $SU(3).$  Let $\psi$ be fixed, let $U$ be a unitary transformation
and let $\bar{U}$ be that transformation that maps $\psi$ to
 \[ \bar{U}\psi = \psi_{0} = \left( \begin{array}{c}
\alpha \\ \alpha* \\ 0 \\ 0
\end{array} \right)  \]
Then for every transformation $U$ that leaves $\psi$ invariant,
       \[ U\psi  = \psi, \]
we can construct another one,
 \[ U_{0} = \bar{U}U\bar{U}^{\dagger}, \]
 that leaves $\psi_{0}$ invariant, and vice-versa:
  \[ U_{0}\psi_{0} = (\bar{U}U\bar{U}^{\dagger}) (\bar{U}\psi) = \bar{U}U\psi
= \bar{U}\psi = \psi_{0} \]
  \[ U\psi = (\bar{U}^{\dagger}U_{0}\bar{U}) (\bar{U}^{\dagger}\psi_{0}) =
\bar{U}^{\dagger}U_{0}\psi_{0} = \bar{U}^{\dagger}\psi_{0} = \psi  \]

Therefore, the group that leaves $\psi$ invariant is the same as the group that
leaves $\psi_{0}$ invariant.  But this group was shown by Wheeler \cite{Wh2} to
be $SU(3),$ by a direct construction of the infinitesimal Hermitian generators.
These generators are required to satisfy $H\psi = 0.$

The existence of a one-parameter, bounded symmetry, $C(1),$ follows immediately
from the expression for $[g]$ in terms of $\psi$.  Each component of $\psi$ is
bounded by the norm of $[g].$  Furthermore, $\psi$ has four complex components,
constrained by the constancy of the norm, $\psi^{\dagger}\psi$. Therefore,
seven
degrees of freedom in $\psi$ parameterize the six
 independent components of $[g]$,
leaving a one-parameter family of solutions to the algebraic equations for
$\psi_{\alpha}([g]).$  Since the equations to be solved for $\psi([g])$ are of
quadratic or higher order, there will be more than a single root, providing a
discrete symmetry, $K.$  Additional discrete symmetry may also be provided by
the phase transformations of the components.

The spacetime metric has definite Weyl weight if and only if $[g]$ has only one
nonvanishing component.  In these cases $\psi$ has exactly two nonvanishing
components.  For example, we may have
     \[ \psi = \left( \begin{array}{c} \alpha \\
 \beta \\ 0 \\ 0 \end{array} \right) \]
If the phase of $\alpha$ is shifted by $\delta$ and the phase of $\beta$ by
$-\delta$, $[g]_{12} = Re(\alpha\beta)$ remains invariant.  This is a $U(1)$
symmetry.  Note that it does not commute with the $SU(3)$ symmetry, but for any
change of phase, $\delta$, it is trivial to write down the new generators of
$SU(3).$

Finally, when the metric is of definite Weyl weight, there is an Hermitian
transformation, $A$, which commutes with $SU(3)$ satisfying $A\psi = \psi.$
When exponentiated, the net effect is a phase change
          \[ U\psi = e^{i\delta A} \psi = e^{i\delta} \psi \]
 In general this does not leave $[g]$ invariant.  However, when $\delta = n\pi,
[g]$ is unchanged.  The transformation remains nontrivial when acting on
spinors
other than $\psi$, and is distinct for different values of $n.$  The symmetry
group, $K,$ therefore contains the integers.
               \end{theorem}

\section{The $SU(2)$ symmetry}

There is a remaining symmetry of the metric choice, which is most naturally
thought of as arising because the metric is symmetric in the gauge fields
$e^{i}_{\alpha}{}^{a}$.  It is natural to ask about the character of the
antisymmetric combination
    \[ F^{i}_{\alpha\beta} = \epsilon^{i}_{jk} e^{j}_{\alpha}{}^{a}
             e^{k}_{\beta}{}^{b} \eta_{ab} = - F^{i}_{\beta\alpha} \]
where $\epsilon_{ijk}$ is the Levi-Civita symbol.  $F^{i}_{\alpha\beta}$ has
the
index structure of a Yang-Mills field.  Applying the same criteria used to
arrive at $SU(4),$ we have the space $R^{3} - \{0\}$ covered by $R \times
O(3),$
with compact part $O(3)$ and covering group $SU(2).$  This is precisely the
additional group required to give the standard model.

   For  $F^{i}_{\alpha\beta}$ to be a gauge field, it must arise from a gauge
potential.  Interestingly, the necessary condition depends on the vanishing of
part of the torsion.  Normally, in the gauging of Poincar\'{e} symmetry, we
impose the condition
        \[ D_{[ \alpha}e_{\beta ]}^{a} = T^{a}_{\alpha\beta} = 0  \]
on the vierbein.  It is sufficient to demand that each metric in our class be
torsion-free.  We therefore require the same condition of each of the three
vierbein components,
                \[ D_{[ \alpha}e^{k}_{\beta ]}{}^{a} = 0 \]
Contracting with $\eta_{ab} \epsilon^{i}_{jk} e^{j}_{\mu}{}^{b}$ and
antisymmetrizing yields the Bianchi identity for $F^{i}_{\alpha\beta}:$
         \[ D_{[ \alpha}F^{i}_{\mu\beta ]} = 0 \]
$F^{i}_{\alpha\beta}$ therefore arises from an $SU(2)$ gauge potential,
providing the final symmetry required for the standard model.

\section{Gauging}

 We now have shown the existence of $SU(3) \times SU(2) \times U(1) \times Z$
symmetry as the residual gauge group following any choice of metric of definite
scaling weight.  The full symmetry is therefore the standard model, together
with the Poincar\'{e} group and a discrete symmetry.  We assumed that the
dilation symmetry is broken even though it does not directly give a scaling of
the mass as assumed by Wess \cite{Wess}.  Even if dilations were allowed, the
standard model symmetries would still remain.

It is important to note that the new unitary symmetries are  independent
of the Poincar\'{e} symmetry.  The Poincar\'{e} symmetry is the remnant of the
original conformal symmetry.  The unitary symmetry was introduced to classify
the metrics allowed by the conformal gauge fields, but has no direct
relationship to the translation, rotation or boost symmetries.  The gauging of
the group may proceed along the usual lines, with the exception that the
product
of the electromagnetic and strong symmetries is semi-direct and not direct.

Still more interesting is the possibility of investigating what happens if the
entire $SU(4)$ symmetry is maintained.  The spacetime metric may be regarded as
an $SU(4)$-valued tensor field, $g^{A}_{\alpha\beta}$, and the curvature for
the
full
symmetry derived.  It remains to be seen whether an appropriate action emerges
naturally in this approach.

Finally, we conjecture that the correct way to introduce matter fields is
through supersymmetrization of the model.


\begin{thebibliography}{9}
      \bibitem {Stdmodel} C. N. Yang and R. L. Mills, Phys. Rev. {\bf 96}
(1954)
191. \\ R. Utiyama, Phys. Rev. {\bf 101} (1956) 1597. \\  S. L. Glashow, Nucl.
Phys. {\bf 22} (1961) 579. \\  S. Weinberg, Phys. Rev. Lett. {\bf 19} (1967)
1264. \\
A. Salam, In {\em Elementary Particle Theory,} ed. N. Svartholm, Stockholm,
Almquist Forlag AB, (1968) 367. \\  C. N. Yang, Ann. N. Y. Acad. Sci. {\bf 294}
(1977) 86.
     \bibitem{Wh1} J.T.Wheeler, to be published in Phys.Rev.D.
     \bibitem{Kaku} M. Kaku, P.K. Townsend and P. Van Nieuwenhuizen, Phys.
Lett.
{\bf 69B} (1977) 304.
     \bibitem{Wess} J. Wess, Nuovo Cim.{\bf 18} (1960) 1086.
     \bibitem{Carruthers} P. Carruthers, Phys. Rep. 1, no.1 (1971) 1.
     \bibitem{Wh2} J.T. Wheeler, USU preprint FTG-105-USU, June 1991 (submitted
for publication).
\end{thebibliography}
\end{document}